\documentclass[conference]{IEEEtran}

\usepackage{amsmath}
\usepackage{amssymb}
\usepackage{graphicx}
\usepackage[usenames,dvipsnames,svgnames,table]{xcolor}

\def\l{\left}
\def\r{\right}
\def \tfidf {\mbox{tf-idf}}

\newcommand{\f}{\frac}
\newcommand{\mean}[1]{\l<#1\r>}
\newcommand{\brkdots}{[\ldots]}
\newcommand{\lett}[1]{(\textbf{#1})}

\DeclareMathOperator{\OR}{OR}

\newcommand{\sad}[0]{\cellcolor{red!30}}
\newcommand{\happy}[0]{\cellcolor{yellow!50}}

\date{\today} 

\begin{document}

%\title{What we talk about when we talk about causality: Lexical, emotional, and topical features of causal statements across large-scale social discourse}
%\title{What we talk about when we talk about causality: Features of causal statements across large-scale social discourse}
\title{What we write about when we write about causality: Features of causal statements across large-scale social discourse}

\author{\IEEEauthorblockN{Thomas C.~McAndrew\IEEEauthorrefmark{1}\IEEEauthorrefmark{2},
Joshua C.~Bongard\IEEEauthorrefmark{3}\IEEEauthorrefmark{2},
Christopher M.~Danforth\IEEEauthorrefmark{1}\IEEEauthorrefmark{2}, 
Peter S.~Dodds\IEEEauthorrefmark{1}\IEEEauthorrefmark{2},\\
Paul D.~H.~Hines\IEEEauthorrefmark{4}\IEEEauthorrefmark{2} and
James P.~Bagrow\IEEEauthorrefmark{1}\IEEEauthorrefmark{2}
}
\IEEEauthorblockA{\IEEEauthorrefmark{1}Department of Mathematics and Statistics, University of Vermont, Burlington, VT, United States}
\IEEEauthorblockA{\IEEEauthorrefmark{2}Vermont Complex Systems Center, University of Vermont}
\IEEEauthorblockA{\IEEEauthorrefmark{3}Department of Computer Science, University of Vermont}
\IEEEauthorblockA{\IEEEauthorrefmark{4}School of Engineering, University of Vermont}}

%
%\author[1,2,*]{Thomas C.~McAndrew}
%\author[3,2]{Joshua C.~Bongard}
%\author[1,2]{Christopher M.~Danforth}
%\author[1,2]{Peter S.~Dodds}
%\author[4,2]{Paul D.~H.~Hines}
%\author[1,2]{James P.~Bagrow}
%\affil[1]{Department of Mathematics and Statistics, University of Vermont, Burlington, VT, United States}
%\affil[2]{Vermont Complex Systems Center, University of Vermont}%, Burlington, VT, United States}
%\affil[3]{Department of Computer Science, University of Vermont}
%\affil[4]{School of Engineering, University of Vermont}%, Burlington, VT 05405}
%\affil[*]{\corrauthinfo{Thomas.McAndrew@uvm.edu}{uvm.edu/\textasciitilde tmcandre/}}

\maketitle

\begin{abstract}
Identifying and communicating relationships between causes and effects is important for understanding our world, but is affected by language structure, cognitive and emotional biases, and the properties of the communication medium. Despite the increasing importance of social media, much remains unknown about causal statements made online. To study real-world causal attribution, we extract a large-scale corpus of causal statements made on the Twitter social network platform as well as a comparable random control corpus. We compare causal and control statements using statistical language and sentiment analysis tools. We find that causal statements have a number of significant lexical and grammatical differences compared with controls and tend to be more negative in sentiment than controls. Causal statements made online tend to focus on news and current events, medicine and health, or interpersonal relationships, as shown by topic models. By quantifying the features and potential biases of causality communication, this study improves our understanding of the accuracy of information and opinions found online.
\end{abstract}

\renewcommand\IEEEkeywordsname{Keywords}

\section{Introduction}

Social media and online social networks now provide vast amounts of data on human online
discourse and other activities~\cite{lazer2009life,asur2010predicting,kaplan2010users,pak2010twitter,wu2011says,dodds2011temporal,mitchell2013geography}. With so much communication taking place
online and with social media being capable of hosting powerful misinformation campaigns~\cite{ratkiewicz2011detecting} such as those claiming vaccines cause autism~\cite{salathe2011assessing,larson2011addressing}, it is more important than ever to better understand the
discourse of causality and the interplay between online communication and the
statement of cause and effect.

Causal inference is a crucial way that humans comprehend the world,
and it has been a major focus of philosophy, statistics, mathematics,
psychology, and the cognitive sciences.
Philosophers such as Hume and Kant have long argued whether
causality is a human-centric illusion or the discovery of
a priori truth~\cite{hume2012treatise,kant1998critique}.
Causal inference in science is incredibly important, and
researchers have developed statistical measures such as Granger
causality~\cite{granger1969investigating},  mathematical and probabilistic
frameworks~\cite{rubin2011causal,sekhon2008neyman,frangakis2002principal,pearl2009causality},
and text mining procedures~\cite{girju2002text,pechsiri2007mining,kim2013mining} to better infer causal influence from data.
In the cognitive sciences, the famous perception experiments of
Michotte \emph{et al.} led to a long line of research exploring the
cognitive biases that humans possess when attempting to link cause and
effect~\cite{rolfs2013visual,scholl2000perceptual,joynson1971michotte}.

How humans understand and communicate cause and effect relationships is
complicated, and is influenced by language structure~\cite{kelley1967attribution,taylor1975point,kelley1980attribution,hilton1990conversational}
and sentiment or valence~\cite{bohner1988triggers}.
A key finding is that the perceived emphasis or causal weight changes between the agent (the
grammatical construct responsible for a cause) and the patient (the construct
effected by the cause) depending on the types of verbs used to describe the cause and effect.  Researchers have hypothesized \cite{brown1983psychological} that this is because of the innate
weighting property of the verbs in the English language that humans use to attribute
causes and effects. 
Another finding is the role of a valence bias: the volume and intensity of causal reasoning may increase due to negative feedback or negative events ~\cite{bohner1988triggers}.

Despite these long lines of research, causal attributions made via social media or online social networks have not been well studied.
The goal of this paper is to explore the language and topics of causal
statements in a large corpus of social media taken from Twitter. We hypothesize
that language and sentiment biases play a significant role in these statements,
and that tools from natural language processing and computational linguistics
can be used to study them. We do not attempt to study the factual
correctness of these statements or offer any degree of verification, nor do
we exhaustively identify and extract all causal statements from these data.
Instead, here we focus on statements that are with high certainty causal statements,
with the goal to better understand key characteristics about causal statements
that differ from everyday online communication.

The rest of this paper is organized as follows: In Sec.~\ref{sec:methods}
we discuss our materials and methods, including the dataset we studied, how we
preprocessed that data and extracted a `causal' corpus and a corresponding `control'
corpus, and the details of the statistical and language analysis tools we studied
these corpora with. In Sec.~\ref{sec:results} we present results using these
tools to compare the causal statements to control statements. We conclude with a
discussion in Sec.~\ref{sec:discussion}.

\section{Materials and Methods}
\label{sec:methods}

\subsection*{Dataset, filtering, and corpus selection}

Data was collected from a 10\% uniform sample of Twitter posts made during 2013, specifically the Gardenhose API. Twitter activity
consists of short posts called tweets which are limited to 140 characters. 
Retweets, where users repost a tweet to spread its content, were not considered.
 (The spread of causal statements will be considered in future work.\@)
We considered only English-language tweets for this study.
To avoid cross-language effects, we kept only tweets with a user-reported
language of `English' and, as a second constraint, individual tweets needed to match more
English stopwords than any other language's set of stopwords. Stopwords considered for each language
 were determined using NLTK's database~\cite{bird2006nltk}.
A tweet will be referred to as a `document' for the rest of this work.

All document text was processed the same way.
Punctuation, XML characters, and hyperlinks were removed, as were
Twitter-specific ``at-mentions'' and ``hashtags'' (see also the Appendix). There is useful information
here, but it is either not natural language text, or it is Twitter-specific, or
both.
Documents were broken into individual words (unigrams) on whitespace. Casing
information was retained, as we will use it for our Named Entity
analysis, but otherwise all words were considered lowercase only (see also the Appendix).
Stemming~\cite{lovins1968development} and lemmatization~\cite{plisson2004rule}
were not performed.

\textit{Causal documents} were chosen to contain one occurrence only of the exact unigrams: `caused', `causing', or `causes'.  
The word `cause' was not included due to its use as a popular contraction for `because'.  
One `cause-word' per document restricted the analysis to single relationships between two relata.
Documents that contain \emph{bidirectional} words (`associate', `relate',
`connect', `correlate', and any of their stems) were also not selected for
analysis. This is because our focus is on causality, an inherently one-sided
relationship between two objects.
We also did not consider additional synonyms of these cause words, although that could be pursued for future work.
\textit{Control documents} were also selected. These documents did not contain
any of `caused', `causing', or `causes', nor any bidirectional words, and
are further matched temporally to obtain the same number of control documents as causal documents
 in each fifteen-minute period during 2013. 
Control documents were otherwise selected randomly; causal synonyms may be present.
The end result of this procedure identified 965,560 causal and 965,560 control
documents.  Each of the three ``cause-words'', `caused', `causes', and `causing'
appeared in 38.2\%, 35.0\%, and 26.8\% of causal documents, respectively.

\subsection*{Tagging and corpus comparison}

Documents were further studied by annotating their unigrams with
\textbf{Parts-of-Speech} (POS) and \textbf{Named Entities} (NE) tags.
POS tagging was done using NLTK v3.1~\cite{bird2006nltk} which implements an averaged perceptron classifier~\cite{tax2000combining}
trained on the Brown Corpus \cite{francis1979brown}.
(POS tagging is affected by punctuation; we show in the Appendix that our results are  relatively robust to the removal of punctuation.)
POS tags denote the nouns, verbs, and other grammatical constructs present in a document.
Named Entity Recognition (NER) was performed using the 4-class, distributional similarity tagger provided as part of the Stanford CoreNLP v3.6.0
toolkit~\cite{manning2014stanford}.  NER aims to identify and classify proper
words in a text.  The NE classifications considered were: Organization,
Location, Person, and Misc.
The Stanford NER tagger uses a conditional random field
model~\cite{finkel2005incorporating} trained on diverse sets of manually-tagged English-language data (CoNLL-2003)~\cite{manning2014stanford}. Conditional random fields allow dependencies between words
 so that `New York' and `New York Times', for example, are
classified separately as a location and organization, respectively.
These taggers are commonly used and often provide reasonably accurate results,
but there is always potential ambiguity in written text and improving upon these
methods remains an active area of research.

\paragraph*{Comparing corpora}
Unigrams, POS, and NEs were compared between the cause and control corpora using
\textbf{odds ratios} (ORs):
\begin{align}
    \OR(x) = \frac{p_C(x)/ (1-p_C(x))}{p_N(x) / (1-p_N(x))},
\end{align}
where $p_C(x)$ and $p_N(x)$  are the probabilities that a unigram, POS, or NE $x$
occurs in the causal and control corpus, respectively. These probabilities were
computed for each corpus separately as $p(x) = f(x) / \sum_{x' \in V} f(x')$,
where $f(x)$ is the total number of occurrences of $x$ in the corpus and $V$ is
the relevant set of unigrams, POS, or NEs.
Confidence intervals for the ORs were computed using Wald's
methodology~\cite{agresti2011categorical}.

As there are many unique unigrams in the text, when computing unigram
ORs we focused on the most meaningful unigrams within each corpus by using the following
filtering criteria:
we considered only the $\OR$s of the 1500 most frequent unigrams in that corpus that also have
a term-frequency-inverse-document-frequency (tf-idf) score above the 90th percentile for that corpus~\cite{manning2003natural}.
The tf-idf was computed as
\begin{equation}
  \tfidf(w) = \log f(w) \times \log \l(\f{D}{\mathit{df}(w)} \r) \label{tfidf},
\end{equation}
where $D$ is the total number of documents in the corpus, and $\mathit{df}(w)$ is the number of
documents in the corpus containing unigram $w$.  Intuitively, unigrams with higher tf-idf
scores appear frequently, but are not so frequent that they are ubiquitous through
all documents. Filtering via tf-idf is standard practice in the
information retrieval and data mining fields.

\subsection*{Cause-trees}

For a better understanding of the higher-order language structure present in
text phrases, \emph{cause-trees} were constructed.  A cause-tree starts with a root
cause word (either `caused', `causing' or `causes'), then the two most probable
words following (preceding) the root are identified. Next, the root word plus
one of the top probable words is combined into a bigram and the top two most
probable words following (preceding) this bigram are found.  Repeatedly applying
this process builds a binary tree representing the $n$-grams that begin with
(terminate at) the root word.  This process can continue until a certain $n$-gram
length is reached or until there are no more documents long enough to search.

\subsection*{Sentiment analysis}

Sentimental analysis was applied to estimate the emotional content of documents.
Two levels of analysis were used: a method where individual unigrams were given
crowdsourced numeric sentiment scores, and a second method involving a trained
classifier that can incorporate document-level
phrase information.

For the first sentiment analysis, each unigram $w$ was assigned a crowdsourced
``labMT'' sentiment score $s(w)$~\cite{dodds2011temporal}. 
(Unlike \cite{dodds2011temporal}, scores were recentered by subtracting the mean, $s(w) \leftarrow s(w)-\mean{s}$.)
Unigrams determined
by volunteer raters to have a negative emotional sentiment (`hate',`death', etc.)
have $s(w) < 0$, while unigrams determined to have a positive emotional
sentiment (`love', `happy', etc.) tend to have $s(w) > 0$.
Unigrams that have labMT scores and are above the 90th percentile of tf-idf
for the corpus form the set $\tilde{V}$. (Unigrams in $\tilde{V}$
need not be among the 1500 most frequent unigrams.) 
The set $\tilde{V}$ captures 87.9\% (91.5\%) of total unigrams in the causal (control)
corpus.
Crucially, the tf-idf filtering ensures that the words `caused', `causes', and `causing',
which have a slight negative sentiment, are not included and do not introduce a
systematic bias when comparing the two corpora.

This sentiment measure works on a per-unigram basis, and is therefore best suited for large
bodies of text, not short documents~\cite{dodds2011temporal}.
Instead of considering individual documents, the distributions of labMT scores
over all unigrams for each corpus was used to compare the corpora. In addition,
a \textbf{single sentiment score} for each corpus was computed as the average
sentiment score over all unigrams in that corpus, weighed by unigram frequency: 
  $\sum_{w \in \tilde{V}} {f(w) s(w)} \Big/ \sum_{w' \in \tilde{V}} f(w')$.

To supplement this sentiment analysis method, we applied a second method capable of estimating with
reasonable accuracy the sentiment of
individual documents. We used the sentiment classifier~\cite{socher2013recursive} included in the
Stanford CoreNLP v3.6.0 toolkit to documents in each corpus.
Documents were individually classified into one of five categories: very negative, negative,
neutral, positive, very positive. The data used to train this classifier is taken
from positive and negative reviews of movies (Stanford Sentiment Treebank v1.0)~\cite{socher2013recursive}.

\subsection*{Topic modeling}

Lastly, we applied topic modeling to the causal corpus to determine what are the
topical foci most discussed in causal statements.
Topics were built from the causal corpus using Latent Dirichlet Allocation
(LDA)~\cite{blei2003latent}.
Under LDA each document is modeled as a bag-of-words or unordered collection of
unigrams.
Topics are considered as mixtures of unigrams by estimating conditional distributions over unigrams: $P(w|T)$, the
probability of unigram $w$ given topic $T$ and documents are considered as mixtures
of topics via $P(T|d)$, the probability of topic $T$
given document $d$. These distributions are then found via statistical
inference given the observed distributions of unigrams across documents. The
total number of topics is a parameter chosen by the practitioner.
For this study we used the MALLET v2.0.8RC3 topic modeling toolkit
\cite{mccallum2002mallet} for model inference. By inspecting the most probable
unigrams per topic (according to $P(w|T)$), we found 10 topics provided
meaningful and distinct topics.

\section{Results}
\label{sec:results}

\begin{figure*}[ht!]
   \centering
   {\includegraphics[width=0.8\textwidth,trim=10 10 0 2,clip=true]{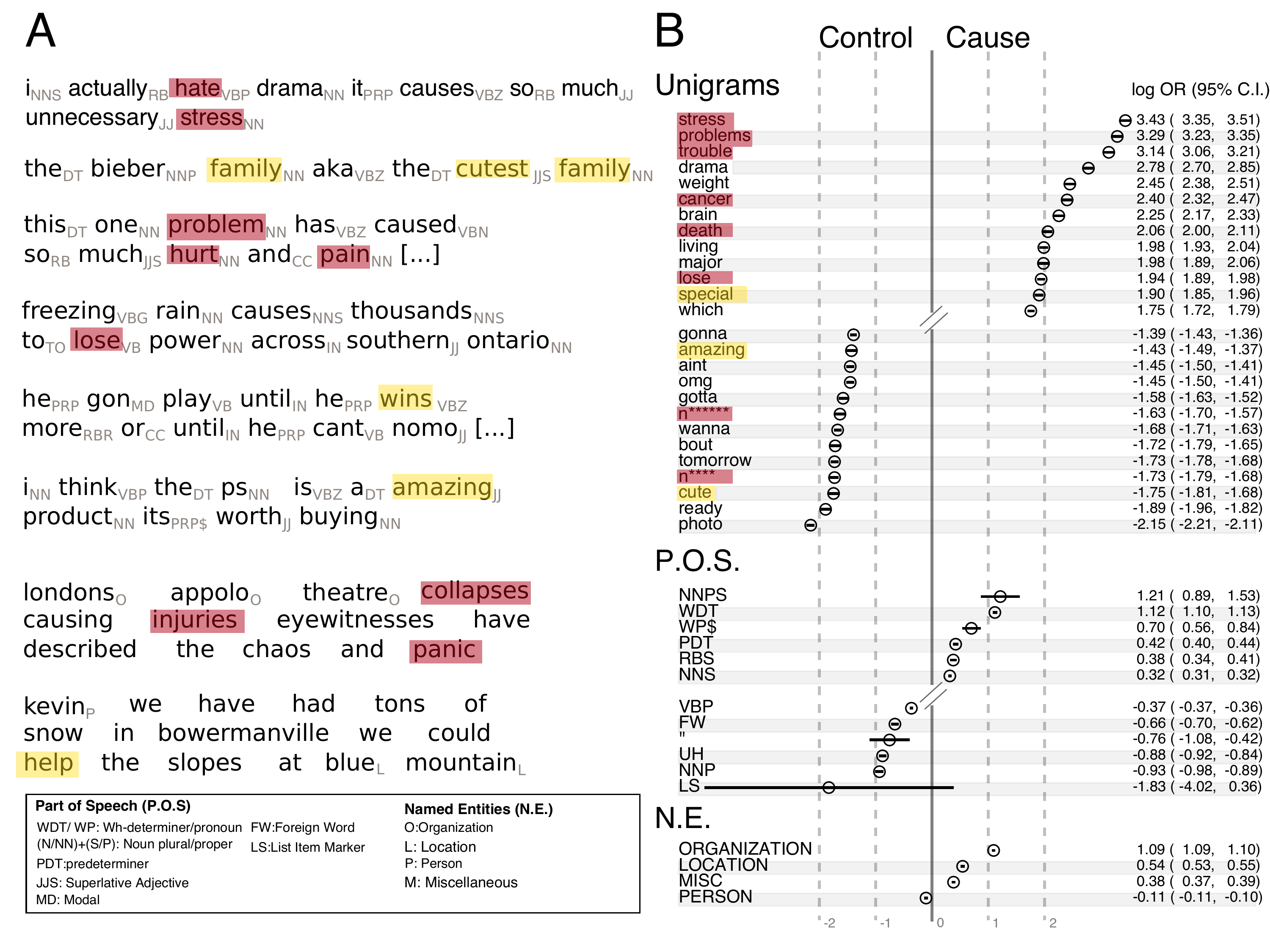}}
   \caption{Measuring the differences between causal and control documents.
   \lett{A} Examples of processed documents tagged by Parts-of-Speech (POS) or Named Entities (NEs). Unigrams highlighted in red (yellow) are in the bottom 10\% (top 10\%) of the labMT sentiment scores. 
   \lett{B} Log Odds ratios with 95\% Wald confidence intervals for the most heavily skewed unigrams, POS, and all NEs between the causal and control corpus. POS tags that are plural and use Wh-pronouns (that, what, which, ...) are more common in the causal corpus, while singular nouns and list items are more common in the controls. Finally, the `Person' tag is the only NE less likely in the causal corpus. 
   Certain unigrams were censored for presentation only, not analysis. All shown odds ratios were significant at the $\alpha = 0.05$ level except LS (List item markers). See also the Appendix.
   \label{fig1.forestplot}}
\end{figure*}

We have collected approximately 1M causal statements made on Twitter over the course of 2013, and for a control we gathered the same number of statements selected at random but controlling for time of year (see Methods). We applied \textbf{Parts-of-Speech} (POS) and \textbf{Named Entity} (NE) taggers to all these texts.
Some post-processed and tagged example documents, both causal and control, are shown in Fig.~\ref{fig1.forestplot}A. We also applied sentiment analysis methods to these documents (Methods) and we have highlighted very positive and very negative words throughout Fig.~\ref{fig1.forestplot}.

In Fig.~\ref{fig1.forestplot}B we present odds ratios for how frequently unigrams (words), POS, or NE appear in causal documents relative to control documents.
The three unigrams most strongly skewed towards causal documents were `stress', `problems', and `trouble', while the three most skewed towards control documents were `photo', `ready', and `cute'. While these are only a small number of the unigrams present, this does imply a negative sentiment bias among causal statements (we return to this point shortly).

%POS tag differences
Figure~\ref{fig1.forestplot}B also presents odds ratios for POS tags, to help us measure the differences in grammatical structure between causal and control documents (see also the Appendix for the effects of punctuation and casing on these odds ratios).
The causal corpus showed greater odds for plural nouns (Penn Treebank tag: NNS), plural proper nouns (NNPS), Wh-determiners/pronouns (WDT, WP\$) such as `whichever',`whatever', `whose', or `whosever', and predeterminers (PDT) such as `all' or `both'.
Predeterminers quantify noun phrases such as `all' in `after \emph{all} the events that caused you tears', showing that many causal statements, despite the potential brevity of social media, can encompass or delineate classes of agents and/or patients.
On the other hand, the causal corpus has lower odds than the control corpus for list items (LS), proper singular nouns (NNP), and interjections (UH).

Lastly, Fig.~\ref{fig1.forestplot}B contains odds ratios for NE tags, allowing us to quantify the types of proper nouns that are more or less likely to appear in causal statements. Of the four tags, only the ``Person'' tag is less likely in the causal corpus than the control. (This matches the odds ratio for the proper singular noun discussed above.)
Perhaps surprisingly, these results together imply that causal statements are less likely to involve individual persons than non-causal statements. There is considerable celebrity news and gossip on social media~\cite{wu2011says}; discussions of celebrities may not be especially focused on attributing causes to these celebrities.
All other NE tags, Organization, Location, and Miscellaneous, occur more frequently in the causal corpus than the control.
All the odds ratios in Fig.~\ref{fig1.forestplot}B were significant at the $\alpha = 0.05$ level except the List item marker (LS) POS tag.

\begin{figure*}[ht!]
   \centering
   {\includegraphics[width=0.85\textwidth,trim=0 6 0 13 0,clip=true]{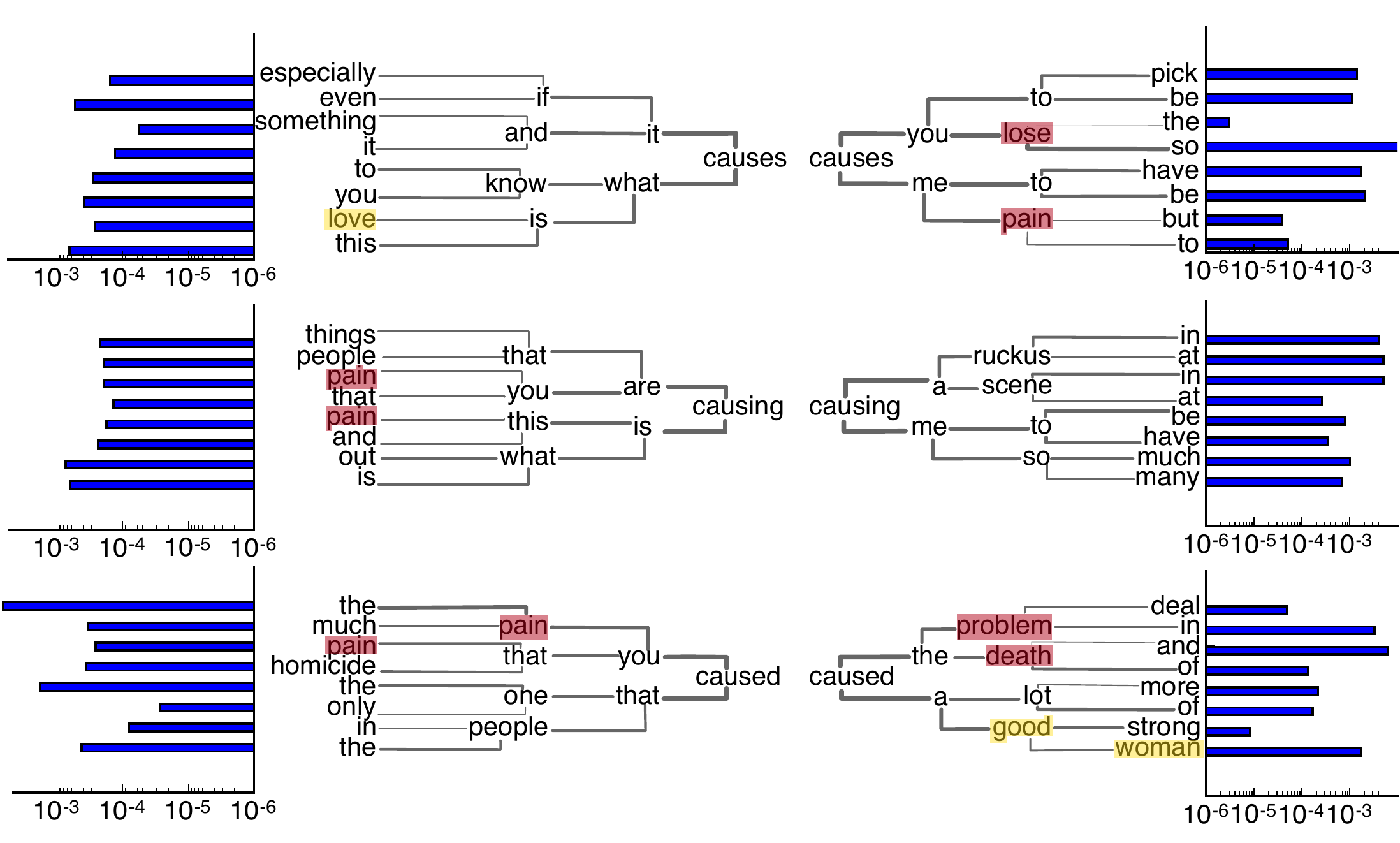}}
   \caption{``Cause-trees'' containing the most probable $n$-grams terminating at (left) or beginning with (right) a chosen root cause-word (see Methods).
   Line widths are log proportional to their corresponding $n$-gram frequency and bar plots measure the 4-gram per-document rate $ N(\mbox{4-gram}) / D$. 
   Most trees express negative sentiment consistent with the unigram analysis (Fig.~\ref{fig1.forestplot}). The `causes' tree shows (i) people think in terms of causal probability (``you know what causes \brkdots{}''), and (ii) people use causal language when they are directly affected or being affected by another (``causes you'', ``causes me''). The `causing' tree is more global (``causing a ruckus/scene'') and ego-centric (``pain you are causing''). The `caused' tree focuses on negative sentiment and alludes to humans retaining negative causal thoughts in the past. 
   \label{fig2.causetrees}}
\end{figure*}

% cause trees
The unigram analysis in Fig.~\ref{fig1.forestplot} does not incorporate higher-order phrase structure present in written language. To explore these structures specifically in the causal corpus, we constructed ``cause-trees'', shown in Fig.~\ref{fig2.causetrees}. Inspired by association mining~\cite{agrawal1993mining},  a cause-tree is a binary tree rooted at either `caused', `causes', or `causing', that illustrates the most frequently occurring $n$-grams that either begin or end with that root cause word (see Methods for details).

The ``causes'' tree shows the focused writing (sentence segments) that many people use to express either the relationship between their own actions and a cause-and-effect (``even if it causes''), or the uncontrollable effect a cause may have on themselves:
 ``causes me to have'' shows a person's inability to control a causal event (``\brkdots{} i have central heterochromia which causes me to have dual colors in both eyes'').
The `causing' tree reveals our ability to confine causal patterns to specific areas, and also our ability to be affected by others causal decisions.
Phrases like ``causing a scene in/at'' and ``causing a ruckus in/at'' (from documents like ``causing a ruckus in the hotel lobby typical \brkdots{}'') show people commonly associate bounds on where causal actions take place.
The causing tree also shows people's tendency to emphasize current negativity: Phrases like ``pain this is causing'' coming from documents like ``cant you see the pain you are causing her'' supports the sentiment bias that causal attribution is more likely for negative cause-effect associations.
Finally, the `caused' tree focuses heavily on negative events and indicates people are more likely to remember negative causal events.
Documents with phrases from the caused tree (``\brkdots{} appalling tragedy \brkdots{} that caused the death'', ``\brkdots{} live with this pain that you caused when i was so young \brkdots{}'') exemplify the negative events that are focused on are large-scale tragedies or very personal negative events in one's life.

Taken together, the popularity of negative sentiment unigrams (Fig.~\ref{fig1.forestplot}) and $n$-grams (Fig.~\ref{fig2.causetrees}) among causal documents shows that emotional sentiment or ``valence'' may play a role in how people perform causal attribution~\cite{bohner1988triggers}. The ``if it bleeds, it leads'' mentality among news media, where violent and negative news are more heavily reported, may appeal to this innate causal association mechanism. (On the other hand, many news media themselves use social media for reporting.) The prevalence of negative sentiment also contrasts with the ``better angels of our nature'' evidence of Pinker~\cite{pinker2011better}, illustrating one bias that shows why many find the results of Ref.~\cite{pinker2011better} surprising. 

\begin{figure*}[ht!]
   \centering
   {\includegraphics[width=0.65\textwidth,trim=0 5 0 10,clip=true]{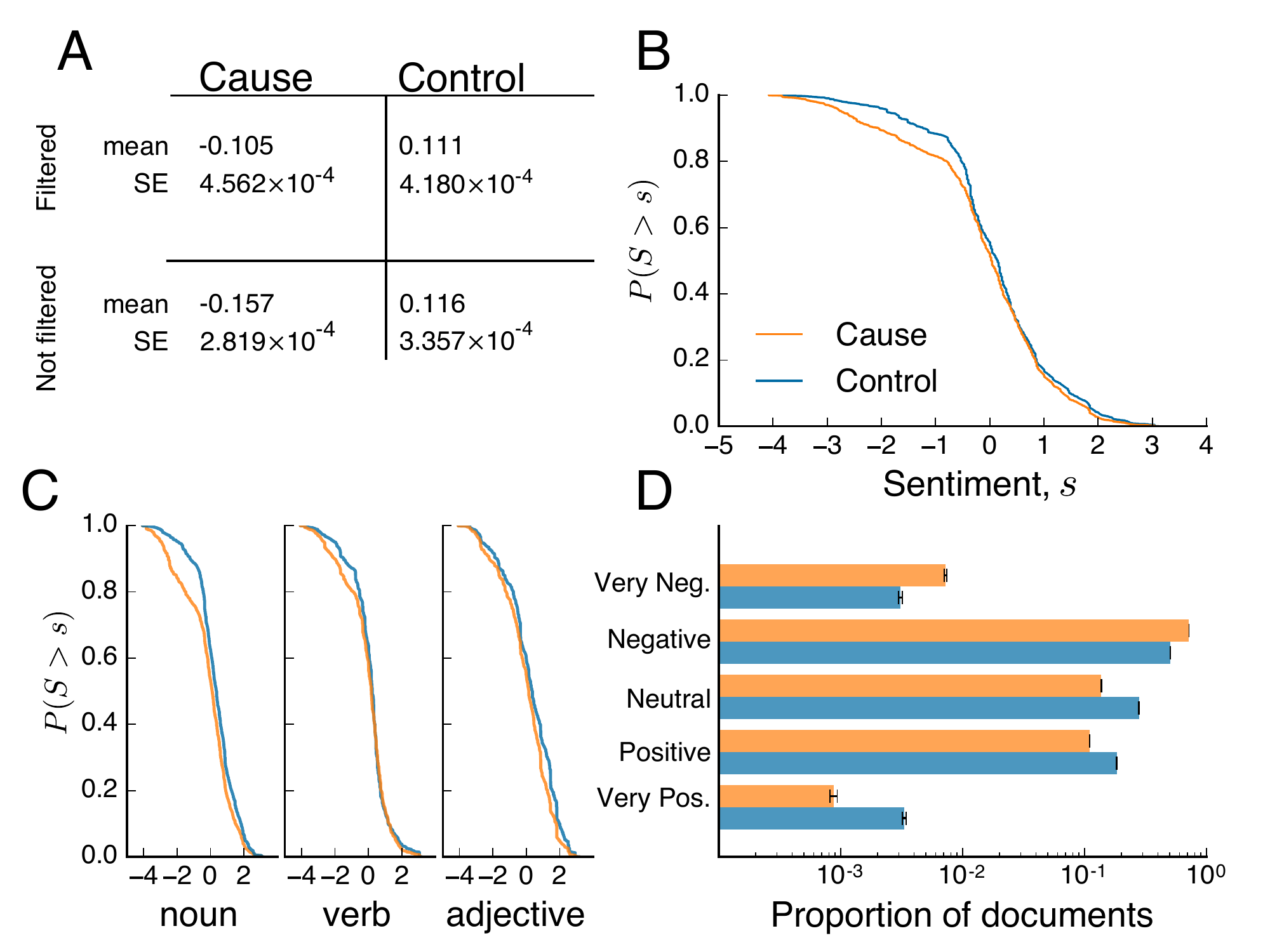}}
   \caption{Sentiment analysis revealed differences between the causal and control corpora. 
   \lett{A} The mean unigram sentiment score (see Methods), computed from crowdsourced ``labMT'' scores~\cite{dodds2011temporal}, was more negative for the causal corpus than for the control. This held whether or not tf-idf filtering was applied.
   \lett{B} The distribution of unigram sentiment scores for the two corpora showed more negative unigrams (with scores in the approximate range $-3 < s < -1/2$) in the causal corpus compared with the control corpus.
   \lett{C} Breaking the sentiment distribution down by Parts-of-Speech, nouns show the most pronounced difference in sentiment between cause and control; verbs and adjectives are also more negative in the causal corpus than the control but with less of a difference than nouns. POS tags corresponding to nouns, verbs, and adjectives together account for 87.8\% and 77.2\% of the causal and control corpus text, respectively. 
   \lett{D} Applying a different sentiment analysis tool---a trained sentiment classifier~\cite{socher2013recursive} that assigns individual documents to one of five categories---the causal corpus had an overabundance of negative sentiment documents and fewer positive sentiment documents than the control. This shift from very positive to very negative documents further supports the tendency for causal statements to be negative.
   \label{fig:sent}}
\end{figure*}

Given this apparent sentiment skew, we further studied sentiment (Fig.~\ref{fig:sent}).
We compared the sentiment between the corpora in four different ways 
to investigate the observation (Figs.~\ref{fig1.forestplot}B and \ref{fig2.causetrees})  that people focus more about negative concepts when they discuss causality.
First, we computed the mean sentiment score of each corpus using crowdsourced ``labMT'' scores weighted by unigram frequency (see Methods).
We also applied tf-idf filtering (Methods) to exclude very common words, including the three cause-words, from the mean sentiment score.
The causal corpus text was slightly negative on average while the control corpus was slightly positive (Fig.~\ref{fig:sent}A).
The difference in mean sentiment score was significant (t-test: $p < 0.01$).

%SENTIMENT

Second, we moved from the mean score to the distribution of sentiment across all (scored) unigrams in the causal and control corpora (Fig.~\ref{fig:sent}B).
The causal corpus contained a large group of negative sentiment unigrams, with labMT scores in the  approximate range $-3 < s < -1/2$; the control corpus had significantly fewer unigrams in this score range.

Third, in Fig.~\ref{fig:sent}C we used POS tags to categorize scored unigrams into nouns, verbs, and adjectives.
Studying the distributions for each, we found that nouns explain much of the overall difference observed in Fig.~\ref{fig:sent}B, with verbs showing a similar but smaller difference between the two corpora. Adjectives showed little difference. The distributions in Fig.~\ref{fig:sent}C account for 87.8\% of 
scored text in the causal corpus and 77.2\% of the control corpus. 
The difference in sentiment between corpora was significant for all distributions (t-test: $p < 0.01$).

Fourth, to further confirm that the causal documents tend toward negative sentiment, we applied a separate, independent sentiment analysis using the Stanford NLP sentiment toolkit~\cite{socher2013recursive} to classify the sentiment of individual documents not unigrams (see Methods). 
Instead of a numeric sentiment score, this classifier assigns documents to one of five categories ranging from very negative to very positive.
The classifier showed that the causal corpus contains more negative and very negative documents than the control corpus, while the control corpus contains more neutral, positive, and very positive documents (Fig.~\ref{fig:sent}D).
%

%Topics

We have found  language (Figs.~\ref{fig1.forestplot} and \ref{fig2.causetrees}) and sentiment  (Fig.~\ref{fig:sent}) differences between causal statements made on social media compared with other social media statements. But \emph{what} is being discussed? What are the topical foci of causal statements? To study this, for our last analysis we applied topic models to the causal statements. Topic modeling finds groups of related terms (unigrams)  by considering similarities between how those terms co-occur across a set of documents.

We used the  popular topic modeling method Latent Dirichlet Allocation (LDA)~\cite{blei2003latent}. We ranked unigrams by how strongly associated they were with the topic. 
Inspecting these unigrams we found that a 10-topic model discovered meaningful topics. See Methods for full details. 
The top unigrams for each topic are shown in Tab.~\ref{tab:topics}.

Topics in the causal corpus tend to fall into three main categories: (i) news, covering current events, weather, etc.; (ii) medicine and health, covering cancer, obesity, stress, etc.;  and (iii) relationships, covering problems, stress, crisis, drama, sorry, etc.

While the topics are quite different, they are all similar in their use of negative sentiment words. 
The negative/global features in the `news' topic are captured in the most representative words: damage, fire, power, etc.
Similar to news, the `accident' topic balances the more frequent day-to-day minor frustrations with the less frequent but more severe impacts of car accidents.
The words `traffic' and `delays' are the most probable words for this topic, and are common, low-impact occurrences. 
On the contrary, `crash', `car', `accident' and `death' are the next most probable words for the accident topic, and generally show a focus on less-common but higher-impact events.

% medicine, body, injuries
The `medical' topic  also focused on negative words; highly probable words for this topic included `cancer', `break', `disease', `blood', etc.
Meanwhile, the `body' topic contained words like: `stress', `lose', and `weight', giving a focus on on our more personal struggles with body image. Besides body image, the `injuries' topic uses specific pronouns (`his', `him', `her') in references to a person's own injuries or the injuries of others such as athletes.

Aside from more factual information, social information is well represented in causal statements.
The `problems' topic shows people attribute their problems to many others with terms like: `dont', `people', `they', `them'. 
The `stress' topic also uses general words such as `more', `than', or `people' to link stress to all people, and in the same vein, the `crisis' topic focuses on problems within organizations such as governments.
The `drama' and `sorry' topics tend towards more specific causal statements.
Drama used the words: `like', `she', and `her'  while documents in the sorry topic tended to address other people.  

The topics of causal documents discovered by LDA showed that both general and specific statements are made regarding news, medicine, and relationships when individuals make causal attributions online.

\begin{table*}%[ht!]
 {\scriptsize
  \centerline{\begin{tabular}{cccccccccc}
     {\footnotesize ``News''}& 
     {\footnotesize ``Accident''}&
     {\footnotesize ``Problems''}&
     {\footnotesize ``Medical''}&
     {\footnotesize ``Crisis''}&
     {\footnotesize ``Sorry''}& 
     {\footnotesize ``Stress''}&
     {\footnotesize ``Body''}&
     {\footnotesize ``Drama''}& 
     {\footnotesize ``Injuries''}\\
\hline 
\sad damage &  traffic &  dont &  \sad cancer &  their &  any &  more &  \sad stress &  \happy like &  his\\
fire &  delays &  people &  break &  our &  never &  than &  \sad lose &  she &  him\\
power &  \sad crash &  they &  some &  from &  been &  being &  weight &  her &  out\\
via &  car &  \sad problems &  men &  how &  sorry &  over &  stuff &  lol &  back\\
new &  \sad accident &  why &  can &  about &  there &  person &  living &  out &  her\\
news &  \sad death &  about &  \sad disease &  social &  know &  \happy sleep &  quickly &  \sad trouble &  when\\
from &  between &  when &  from &  \sad crisis &  will &  which &  \happy special &  \happy good &  head\\
says &  after &  know &  most &  via &  ive &  people &  proof &  now &  into\\
after &  year &  them &  our &  \happy great &  they &  one &  diets &  \sad sh*t &  off\\
video &  down &  \happy like &  others &  \happy money &  out &  \sad stress &  excercise &  \happy life &  well\\
global &  there &  drama &  \sad loss &  many &  \sad problems &  someone &  f*ck &  twitter &  which\\
rain &  man &  who &  \happy heart &  issues &  can &  makes &  who &  got &  from\\
warming &  due &  one &  \happy health &  should &  now &  think &  giving &  scene &  down\\
water &  from &  youre &  \happy food &  \sad war &  \sad trouble &  when &  \happy love &  get &  game\\
\sad explosion &  snow &  get &  symptoms &  \sad problems &  see &  most &  \happy god &  too &  fall\\
outage &  road &  stop &  hair &  government &  one &  thinking &  people &  girl &  face\\
storm &  old &  think &  \happy women &  true &  how &  brain &  will &  \happy haha &  then\\
change &  over &  how &  blood &  new &  would &  \sad depression &  our &  needs &  get\\
house &  \sad problems &  \sad sh*t &  how &  world &  could &  \sad anxiety &  those &  see &  \sad injuries\\
may &  chaos &  want &  skin &  they &  were &  \sad lack &  one &  walk &  had\\
flooding &  morning &  cant &  records &  media &  had &  night &  around &  drama &  sports\\
gas &  two &  because &  adversity &  other &  ever &  without &  \happy life &  hes &  stick\\
air &  driving &  too &  high &  obama &  whats &  \happy love &  his &  woman &  over\\
say &  major &  \sad hate &  helium &  financial &  again &  mental &  thats &  some &  while\\
stir &  today &  need &  which &  change &  did &  them &  work &  last &  eyes\\
heavy &  disruption &  only &  \happy eating &  \sad violence &  time &  mind &  out &  strong &  only\\
weather &  train &  really &  may &  will &  think &  fact &  \happy good &  shes &  hit\\
\sad collapse &  accidents &  many &  body &  also &  well &  insomnia &  \happy sex &  always &  famous\\
climate &  almost &  even &  \sad smoking &  shutdown &  something &  hand &  come &  him &  hockey\\
\sad death &  into &  then &  own &  issue &  \sad ill &  even &  \happy great &  ways &  right\\
\sad deaths &  driver &  someone &  acne &  support &  still &  feel &  say &  little &  left\\
\happy home &  police &  their &  \sad death &  \happy kids &  about &  physical &  back &  because &  \sad injury\\
oil &  until &  away &  brain &  \sad problem &  sure &  emotional &  when &  said &  got\\
massive &  delay &  always &  alcohol &  \sad poor &  \happy hope &  become &  give &  really &  room\\
\sad attack &  congestion &  feel &  common &  \happy free &  get &  can &  their &  thats &  involvement\\
blast &  school &  thats &  \sad deaths &  says &  youve &  too &  them &  here &  innumerable\\
two &  late &  say &  news &  pay &  thats &  less &  things &  man &  time\\
city &  weather &  thing &  treatment &  against &  day &  same &  goes &  ass &  \happy play\\
into &  been &  something &  unknown &  \happy party &  some &  often &  comes &  night &  run\\
state &  earlier &  yourself &  \sad damage &  confusion &  \happy good &  keeps &  too &  ego &  because\\
 \end{tabular}}}
\caption{Topical foci of causal documents. Each column lists the unigrams most highly associated (in descending order) with a topic, computed from a 10-topic Latent Dirichlet Allocation model. The topics generally fall into three broad categories: news, medicine, and relationships. 
Many topics place an emphasis on negative sentiment terms.
Topic names were determined manually. Words are highlighted according to sentiment score as in Fig.~\ref{fig1.forestplot}. 
\label{tab:topics}}
\end{table*}

\section{Discussion}
\label{sec:discussion}

The power of online communication is the speed and ease with which information can be propagated by potentially any connected users. Yet these strengths come at a cost: rumors and misinformation also spread easily. Causal misattribution is at the heart of many rumors, conspiracy theories, and misinformation campaigns.
%
%The goal of this analysis was to understand what it is that people talk about when they talk about causal relationships.
%We approached this problem from language analysis (unigram/$n$-gram, parts-of-speech, named entities), sentiment analysis, and topic modeling.
%We found a distinct negative sentiment among causal statements including a focus on topics such as health scares and violent news events. Causal statements were less likely than other social media posts to include persons (lower odds for the named entity `person' tag and for the proper singular noun part-of-speech). 

Given the central role of causal statements, further studies of the interplay of information propagation and online causal attributions are crucial. Are causal statements more likely to spread online and, if so, in which ways? What types of social media users are more or less likely to make causal statements? Will a user be more likely to make a causal statement if they have recently been exposed to one or more causal statements from other users?

The topics of causal statements also bring forth important questions to be addressed: how timely are causal statements? Are certain topics always being discussed in causal statements? Are there causal topics that are very popular for only brief periods and then forgotten? Temporal dynamics of causal statements are also interesting: do time-of-day or time-of-year factors play a role in how causal statements are made? %What about geographic factors? 

Our work here focused on a limited subset of causal statements, but more generally, these results may inform new methods for automatically detecting causal statements from unstructured, natural language text~\cite{girju2002text}. Better computational tools focused on causal statements are an important step towards further understanding misinformation campaigns and other online activities.
Lastly, an important but deeply challenging open question is how, if it is even possible, to validate the \emph{accuracy} of causal statements. Can causal statements be ranked by some confidence metric(s)? We hope to pursue these and other questions in future research.

%\vspace{-0.4375em}

\appendix{}

%\paragraph*{Punctuation, Casing, and Parts-of-Speech}

\begin{figure}[htbp]
   \centering
   {\includegraphics[width=0.9\columnwidth,trim=0 5 0 5,clip=true]{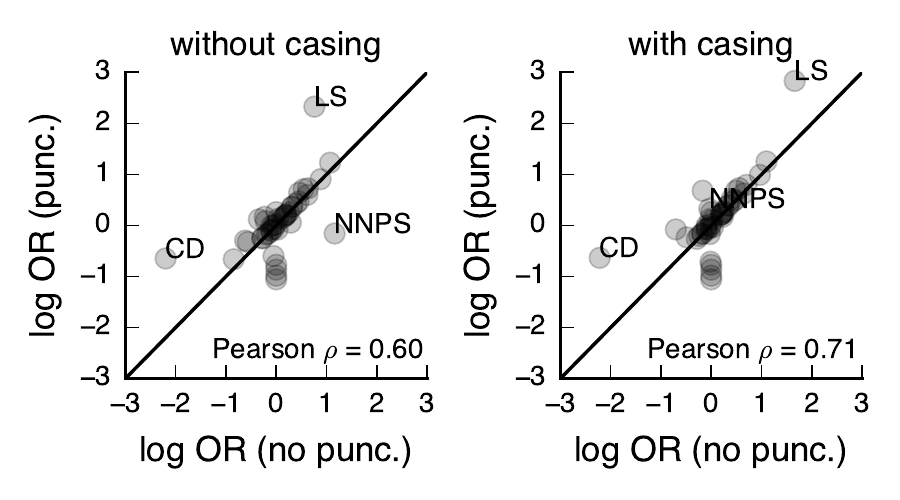}}
\caption{Comparison of Odds Ratios for all Parts-of-Speech (POS) tags with punctuation retained and removed for documents with and without casing. Tags Cardinal number (CD), List item marker (LS), and Proper noun plural (NNPS) were most affected by removing punctuation.
\label{fig:appORscatter}
}
\end{figure}

Parts-of-speech tagging depends on punctuation and casing, which we filtered in our data, so a study of how robust the POS algorithm is to punctuation and casing removal is important.
We computed POS tags for the corpora with and without casing as well as with and without punctuation (which includes hashtags, links and at-symbols).
Two tags mentioned in Fig.~\ref{fig1.forestplot}B, NNPS and LS (which was not significant), were affected by punctuation removal.
Otherwise, there is a strong correlation (Fig.~\ref{fig:appORscatter}) between Odds Ratios (causal vs.\ control) with punctuation and without punctuation, including casing and without casing ($\rho = 0.71$ and $0.80$, respectively), indicating the POS differences between the corpora were primarily not due to the removal of punctuation or casing.

\section*{Acknowledgments}

We thank R.~Gallagher for useful comments and gratefully acknowledge the resources provided by the Vermont Advanced Computing Core.
This material is based upon work supported by the National Science Foundation under Grant No.\ ISS-1447634.

% Generated by IEEEtran.bst, version: 1.13 (2008/09/30)

\end{document}